# Quasi-indirect measurement of electrocaloric temperature change in PbSc$_{0.5}$Ta$_{0.5}$O$_3$ via comparison of adiabatic and isothermal electrical polarization data

S. Crossley[1], R. W. Whatmore[2], N. D. Mathur[1,*] and X. Moya[1,+]

[1]Department of Materials Science, University of Cambridge, Cambridge, CB3 0FS, UK

[2]Department of Materials, Royal School of Mines, South Kensington Campus, Imperial College London, London SW7 2AZ, UK

*ndm12@cam.ac.uk, +xm212@cam.ac.uk

Electrically driven adiabatic changes of temperature are identified in the archetypal electrocaloric material PbSc$_{0.5}$Ta$_{0.5}$O$_3$ by comparing isothermal changes of electrical polarization due to slow variation of electric field, and adiabatic changes of electrical polarization due to fast variation of electric field. By obtaining isothermal (adiabatic) electrical polarization data at measurement (starting) temperatures separated by <0.4 K, we identify a maximum temperature change of ~2 K due to a maximum field change of 26 kV cm$^{-1}$, for starting temperatures in the range 300 - 315 K. These quasi-indirect measurements combine with their direct, indirect and quasi-direct counterparts to complete the set, and could find routine use in future.

Voltage-driven thermal changes known as electrocaloric (EC) effects are maximised near phase transitions in ferroelectric materials,[1,2] and can be used to pump heat in cooling cycles if the process is nominally reversible,[2] such that thermal changes and the concomitant changes of electrical polarization have equal magnitude on field application and field removal despite any field hysteresis. EC effects have now been exploited in a number of prototype cooling devices, where the flow of heat is driven by the temperature change that can be achieved in the working body under adiabatic conditions.[3-16] The EC behaviour of a given material may be identified in terms of an adiabatic temperature change $\Delta T$, an isothermal entropy change $\Delta S$, or the corresponding isothermal heat $Q$. These parameters can be obtained via direct measurements of $\Delta T$ or $Q$, quasi-direct measurements of heat that most traditionally yield $\Delta S$, and indirect measurements that most immediately yield $\Delta S$ from isothermal measurements of electrical polarization $P$ versus electric field $E$ (or $\Delta T$ from adiabatic measurements of $P$).[1-2,17] Noting that these parameters can be interconverted, either crudely via $c|\Delta T| \sim T|\Delta S| = |Q|$ using some effective value of specific heat capacity $c(T,E)$,[1] or more precisely by constructing detailed maps of $S(T,E)$ or $T(S,E)$ using $c(T,0)$ without requiring $c(T,E)$,[17-19] we complete here the set of EC measurement techniques by demonstrating quasi-indirect measurements that yield $|\Delta T|$.

Using a single sample of $PbSc_{0.5}Ta_{0.5}O_3$ (PST), we made fast adiabatic $P_{adi}(E)$ measurements at closely spaced values of zero-field temperature $T_z$, and slow isothermal $P_{iso}(E)$ measurements at closely spaced values of measurement temperature $T$. For analysis, we use the outer branches of $P_{adi}(E)$ and $P_{iso}(E)$ in $E \geq 0$, which are equivalent (similar[17]) to the outer (inner) branches of unipolar cycles that would yield cooling (heating) in EC applications. Note that we performed bipolar rather than unipolar measurements here, as unipolar cycles must be accompanied by bipolar cycles in order to position unipolar plots on the polarization axis, such that the viable acquisition of unipolar cycles would shorten the measurement time for each $P_{iso}(E)$ branch in light of the 30 s measurement constraint discussed later.

Using these outer branches in $E \geq 0$, we identify the nominally reversible adiabatic temperature change between any two points in $P_{adi}(E)$ as the temperature difference between the two $P_{iso}(E)$ plots that intersect these points. Note that the method could be executed just as accurately using the measured $Q(V)$ data, without the geometrical normalization that results in a $P(E)$ dataset. However, the resulting $P(E)$ data are more familiar, and easier to compare with literature, and the comparative nature of the quasi-indirect method ensures the cancellation of any errors associated with the geometrical measurements.

Our method was inspired by an analogous study of magnetocaloric gadolinium,[20] and its application to EC materials, using much denser data, is novel. It is an indirect method given that we do not make direct measurements of electrically driven temperature change, but it is less indirect than the indirect method because we can identify $|\Delta T|$ without heat capacity data, without requiring knowledge of sample geometry, without thermodynamic analysis (not formally valid with hysteresis, not formally valid for relaxors), and without the resultant data processing that can amplify systematic errors and lead to artefacts. We therefore describe our method as quasi-indirect, which completes the set of measurements and avoids the unwelcome possibility of calling it a second type of quasi-direct method. However, our measurements of temperature ($T$ and $T_z$) rather than electrically driven temperature change are reminiscent of the quasi-direct method,[21,22] where one measures thermally driven isofield heat instead of electrically driven heat.

Our PST sample was similar to ~400 µm-thick samples for which we have reported direct and indirect EC measurements[17]. Those samples and the present sample all came from the same master wafer, and they were thinned, mounted and electroded in the same way. The present sample was 330 µm thick and possessed an area of 0.19 cm$^2$. The Pt bottom electrode was ubiquitous. The Pt top electrode fell ~0.5 mm from the sample edges, and its effective area was 0.11 cm$^2$ after incorporating a ~6% increase for fringing fields.[23] A smear of vacuum grease around the top electrode prevented arcing. While slowly varying measurement temperature $T$ (starting temperature $T_z$) with a cryogenic probe that was fabricated in house,[24] we measured the isothermal (adiabatic) electrical polarization using a Radiant Precision Premier II with a Trek high-voltage amplifier, which integrated the displacement current that resulted from the application of a continuous triangular driving waveform of magnitude 26 kV cm$^{-1}$ and period 30 s (period 0.5 s). Sample temperature was not measured directly in order to avoid the thermal mass associated with contact thermometry.

As reported in ref. [17], our PST displays a first-order ferroelectric phase transition at a Curie temperature of $T_C$ ~ 295 K, consistent with a high degree of B-site cation order (~0.80). The field sweep rates for measuring $P_{iso}(E)$ and $P_{adi}(E)$ were identified by comparing one quarter of the cycle period (during which the field varies between zero and its maximum value) with the thermal timescale, which was found to be ~5 s via direct EC measurements of a similarly mounted similar sample (Supplementary Note 4 in ref. [17]), and which could be determined without direct EC measurements by employing ever more extreme periods until there is no change to $P_{iso}(E)$ and $P_{adi}(E)$. On heating slowly at 0.1 K min$^{-1}$, from 280 K through $T_C$ to 320 K, we obtained $P_{iso}(E)$ plots at 112 values of temperature $T$ separated by ~0.36 K, using the slow 30 s driving period to promote good thermalization while the field was varied [Fig. 1(a) shows six examples]. Plotting the outer branches in $E \geq 0$ for all 112 plots of $P_{iso}(E)$ yields Fig. 1(b). On repeating the temperature sweep, we obtained $P_{adi}(E)$ plots at 358 values of zero-field temperature $T_z$ separated by ~0.11 K, using the fast 0.5 s driving period to avoid any significant thermalization [Fig. 1(d) shows six examples]. Plotting the outer branches in $E \geq 0$ for all 358 plots of $P_{adi}(E)$ yields Fig. 1(e). By assuming that measurements of $P_{adi}(E)$ produce a nominally reversible adiabatic temperature change, we are able to assume that the zero-field temperature $T_z$ prior to measurement is equal to the zero-field temperature on the outer branches of interest.

The individual plots of $P_{iso}(E)$ [Fig. 1(a)] and $P_{adi}(E)$ [Fig. 1(d)] evidence ferroelectricity below $T_C$, paraelectricity above $T_C$, and the electrically driven phase transition (double loop[17,25]) near and above $T_C$. The phase transition is hard to discern by eye when viewing the dense maps of polarization [Fig. 1(b,e)], but it can be clearly seen after the maps have been transposed onto $E$-$T$ axes [Fig. 1(c,f)]. The gradient of the phase boundary is $|dE/dT_0| \sim 1$ kV cm$^{-1}$ K$^{-1}$, as expected for PST from the same master wafer[17] (transition temperature $T_0(E)$ equals $T_C$ at $E = 0$).

Our quasi-indirect method can be understood from Fig. 2(a), where we have used white to copy the $T_z = 305$ K plot of $P_{adi}(E)$ from Fig. 1(d), and where we have used green to copy the 305 K and 307 K plots of $P_{iso}(E)$ from Fig. 1(b). Given that the plot of $P_{adi}(E)$ intersects the 305 K plot of $P_{iso}(E)$ at $E = 0$, and the 307 K plot of $P_{iso}(E)$ at our maximum applied field of $E = 26$ kV cm$^{-1}$, we infer that following the $T_z = 305$ K plot of $P_{adi}(E)$ from 0 to $E = 26$ kV cm$^{-1}$ results in a nominally reversible adiabatic temperature change of $|\Delta T| \sim 307$ K - 305 K = 2 K with respect to $T_z = 305$ K.

Our quasi-indirect method can be implemented more generally to evaluate values of $|\Delta T|$ for changes of field and starting temperatures that lie within our windows of isothermal and adiabatic measurement [Fig. 1(b,e)]. In Fig. 2(b), we reproduce from Fig. 1(b) all 112 plots of $P_{iso}(E)$ on a repeating colourscale, whose 2 K period was chosen to match the value of $|\Delta T|$ that we identified above. Overlaid in white, we reproduce from Fig. 1(e) some of the 358 plots of $P_{adi}(E)$. Specifically, we show relatively well-spaced plots of $P_{adi}(E)$ for intermediate values of $T_z$ that differ by ~2 K. Conceptually, one should imagine all 358 plots of $P_{adi}(E)$ to be present, but of course we cannot show them all without masking the colourful $P_{iso}(E)$ plots. These colourful plots represent a temperature map, and they have inspired us to call this paper the 'rainbow paper'.

One can evaluate $|\Delta T|$ by eye when following a white $P_{adi}(E)$ plot between start and end points that lie around the middle of Fig. 2(b), such that each of these points can be identified with two specific $P_{iso}(E)$ plots that are identified by colour. Suppose that a starting field and a starting temperature correspond to a specific point on a $P_{iso}(E)$ plot that is red. One should then identify the intersecting white plot of $P_{adi}(E)$, which might or might not be one of the 358 $P_{adi}(E)$ plots that we happen to show. One follows the intersecting white plot of $P_{adi}(E)$ up to the finishing field, and evaluates $|\Delta T|$ via the colour change associated with the intersecting plots of $P_{iso}(E)$. For example, if one were to follow a white plot from one red contour to the next then one would have completed one period, implying $|\Delta T| = 2$ K. Inspection of both Fig. 2(a) and Fig. 2(b) indicates that a temperature change of this magnitude might just be possible.

Visual evaluation becomes unreliable when the colourful plots of $P_{iso}(E)$ are closely bunched, but values of $|\Delta T|$ can nevertheless be established by computation, and accuracy can be improved by averaging over similar trajectories (Supplementary Note 1). Using this method, we plot $|\Delta T(T_z)|$ (Fig. 3a) for a field change that approximately corresponds to our maximum field change from 0 to 26 kV cm$^{-1}$, and we find a maximum value of $|\Delta T| \sim 1.7 \pm 0.2$ K in 300 K $< T_z <$ 315 K. Using a statistical method (Supplementary Note 2) yielded a similar plot of $|\Delta T(T_z)|$ (Fig. 3b), again with a maximum value of $|\Delta T| \sim 1.7 \pm 0.2$ K in 300 K $< T_z <$ 315 K, but without the spurious large values of $|\Delta T(T_z)|$ at low and high temperatures [grey in Fig. 3(a)].

Our maximum temperature change of ~1.7 ± 0.2 K occurs in a range of temperatures that is similar with respect to our data for PST from the same wafer,[17] but our peak value is slightly smaller than the directly measured value of $|\Delta T| \sim 2.2$ K (which is itself slightly smaller than the indirectly measured value[17]). Let us now consider the origin of this discrepancy. The $P_{adi}(E)$ data were adiabatic because each branch was measured over 0.125 s (one quarter of the 0.5 s period), which is forty times faster than the ~5 s thermal timescale for exponential decay. However, the $P_{iso}(E)$ data were not properly isothermal because each branch was measured over 7.5 s (one quarter of the 30 s period that represents an upper bound imposed by the proprietary software of the Radiant ferroelectric tester), which is only slightly greater than the ~5 s thermal timescale for exponential decay. This discrepancy with respect to isothermal conditions leads to conservative values of $|\Delta T|$, but it is difficult to quantify the error, partly because the first-order phase transition is locally discontinuous, and partly because heat will be exchanged within the thus phase-separated sample. In future work, one should ensure that $P_{iso}(E)$ is isothermal throughout by varying the applied field sufficiently slowly, e.g. using the constant-current method with a sufficiently low current.[18,26-27] Separately, one should ensure that the measurement temperature is swept slowly enough to measure

$P_\text{iso}(E)$ at temperatures whose separation is as small as possible a fraction of $|\Delta T|$.

In summary, the quasi-indirect method involves following a rapidly acquired plot of adiabatic polarization versus field, and identifying the EC temperature change by reading the temperatures of the isothermal plots that it crosses. Although the thermometer in the cryogenic probe is used to identify starting (measurement) temperatures for the adiabatic (isothermal) plots, it does not measure EC temperature change, permitting one to regard the sample as if it were itself a thermometer. In this respect, the quasi-indirect method differs from the direct, indirect and quasi-direct methods, and may provide an attractive alternative in circumstances where these other methods prove challenging. One exciting possibility is to use the quasi-indirect method to measure free-standing EC films. For example, one might hope to measure films that are tens of microns thick at hundreds of kilohertz (a ten-fold thickness reduction with respect to the present sample implies a hundred-fold reduction in thermalization time[24]). One might also use modelling to back out adiabatic limits that cannot be reached in practice.

**Data Availability Statement**

Data available in article.

**Acknowledgements**

We thank P. C. Osbond for fabricating the master sample. We thank EPSRC (UK) for funding via EP/M003752/1. X.M. is grateful for support from the Royal Society via URF\R\180035.

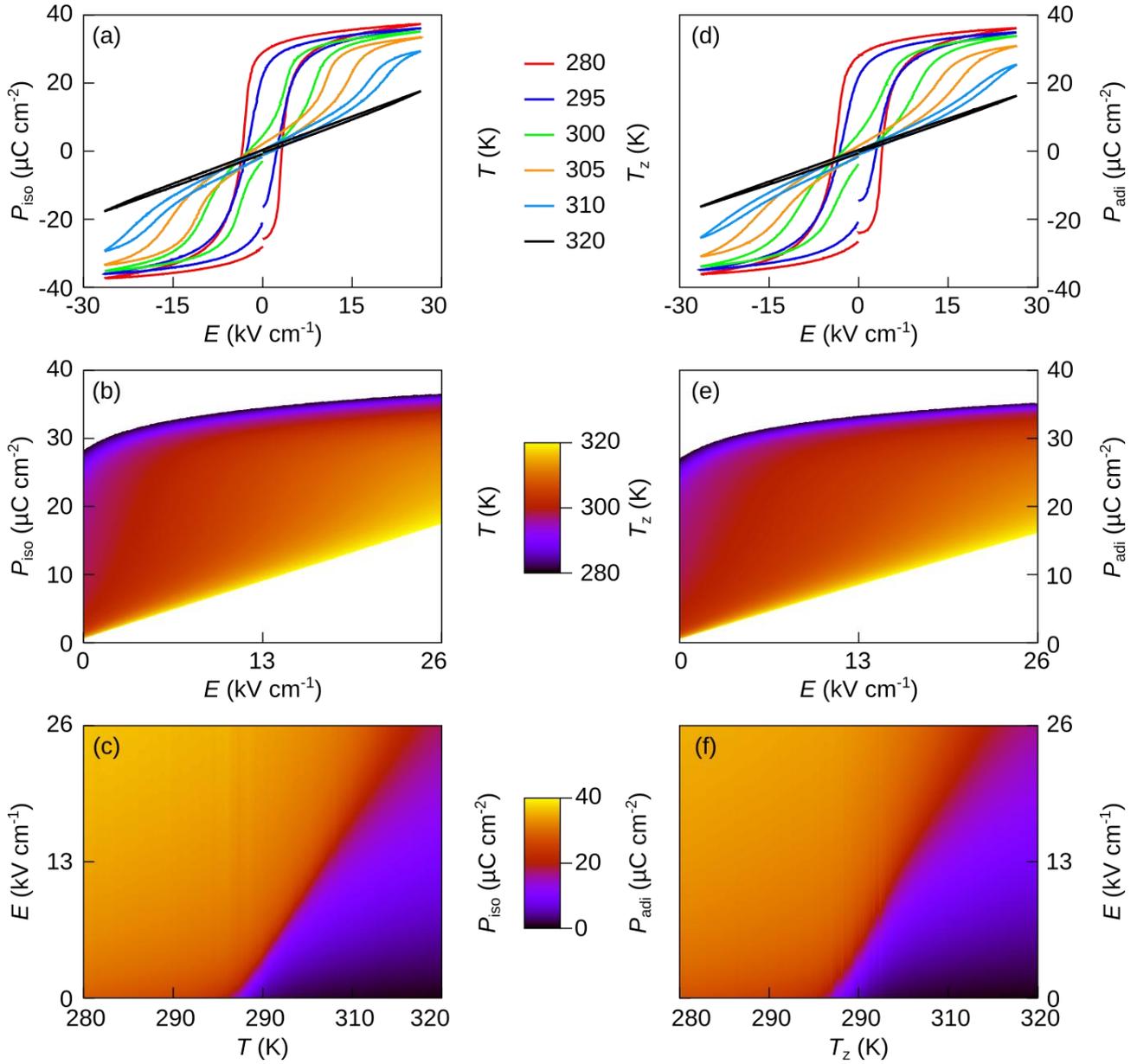

Figure 1. Isothermal and adiabatic measurements of electrical polarization for a single sample of PST. (a-c) Isothermal bipolar $P_{iso}(E)$ plots at 112 values of measurement temperature $T$ every ~0.36 K are presented as (a) $P_{iso}(E)$ for six values of $T$, (b) $T(E,P_{iso})$ constructed from outer branches in $E \geq 0$ for all 112 values of $T$, and hence (c) $P_{iso}(T,E)$. (d-f) Adiabatic bipolar $P_{adi}(E)$ plots at 358 values of zero-field temperature $T_z$ every ~0.11 K are presented as (d) $P_{adi}(E)$ for six values of $T_z$, (e) $T_z(E,P_{adi})$ constructed from outer branches in $E \geq 0$ for all 358 values of $T$, and hence (f) $P_{adi}(E,T_z)$. Values of $T_z$ were set on inner branches at $E = 0$, and represent the temperatures for the outer branches of interest at $E = 0$, assuming nominal reversibility. Note that adiabatic plots outnumber isothermal plots because equivalent warming runs permitted more fast field sweeps, and fewer slow field sweeps.

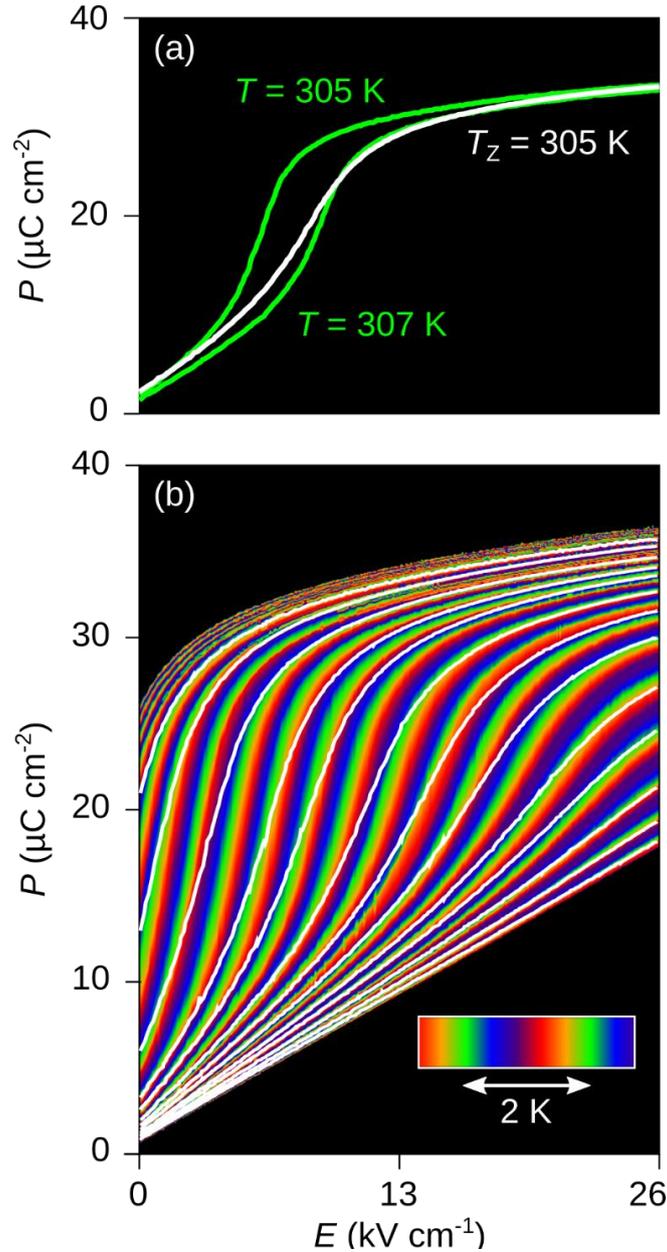

Figure 2. Comparison of the isothermal and adiabatic electrical polarization data in Fig. 1. Data are presented for outer branches in $E \geq 0$. (a) $P_{adi}(E)$ for $T_z = 305$ K (white), and $P_{iso}(E)$ at 305 K (upper green plot) and 307 K (lower green plot). (b) $P_{adi}(E)$ for a subset of $T_z$ values every ~2 K in the middle part of the measurement range such that 280 K $<<$ $T_z$ $<<$ 320 K (white plots), and $P_{iso}(E)$ for all $T$ values every ~0.36 K in 280 K $\leq T_z \leq$ 320 K (periodic colour scale). The temperature for a given $P_{iso}(E)$ plot may be identified by counting repeats of the colour scale with respect to the lowest-lying isotherm at 320 K. Note that all values of $P_{adi}(E)$ were scaled up by 3.1% to match $P_{iso}(E)$ below $T_C$, where EC effects are negligible.[17] This scaling corrects a systematic error in the saturation polarizations that we identified when using a frequency-dependent input impedance to measure very different displacement currents on very different timescales.

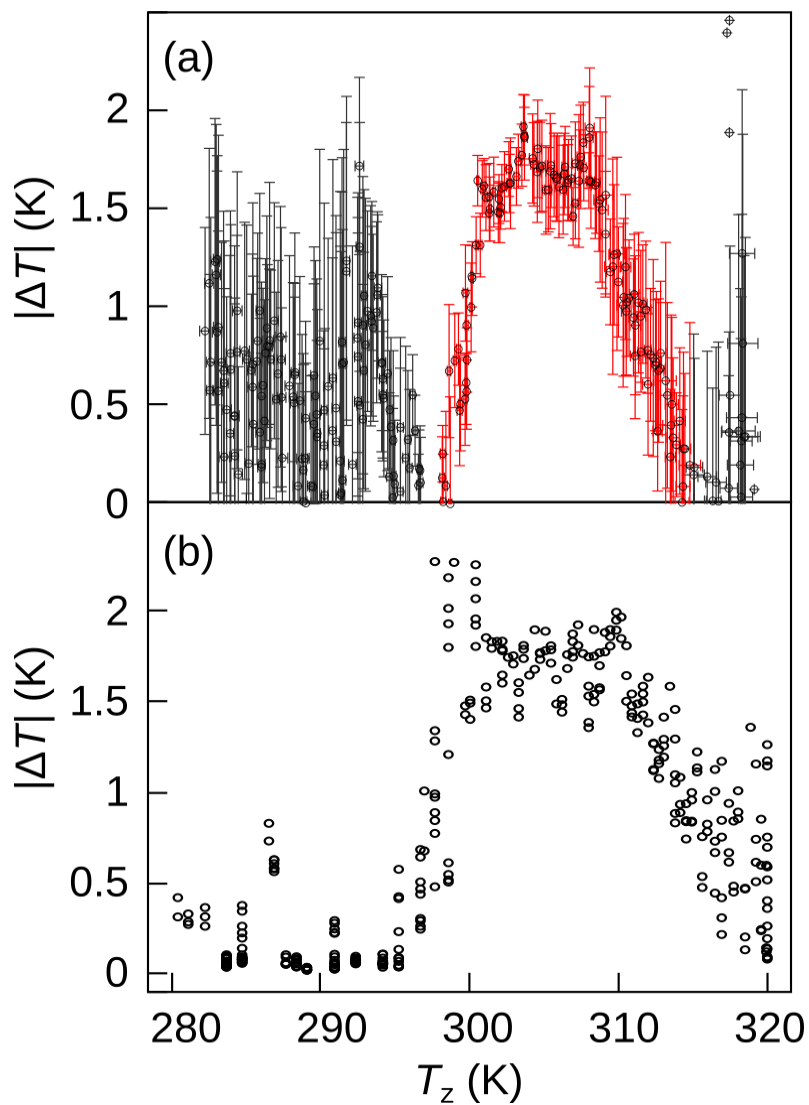

Figure 3. Summary of adiabatic temperature change. Plots of $|\Delta T(T_z)|$ were identified for a change of field between 0 and ~26 kV cm$^{-1}$ using (a) linear interpolation (Supplementary Note 1) and (b) statistical analysis (Supplementary Note 2). Grey data in (a) represent false positive outcomes of the method.

# Supplementary information

# for

# Quasi-indirect measurement of electrocaloric temperature change in PbSc$_{0.5}$Ta$_{0.5}$O$_3$ via comparison of adiabatic and isothermal electrical polarization data

S. Crossley[1], R. W. Whatmore[2], N. D. Mathur[1] and X. Moya[1]

[1]Department of Materials Science, University of Cambridge, Cambridge, CB3 0FS, UK
[2]Department of Materials, Royal School of Mines, South Kensington Campus, Imperial College London, London SW7 2AZ, UK

**Note 1. Evaluation of |Δ$T$| by linear interpolation and averaging**

The map of $T_{iso}(P,E)$ (Fig. 1b) shows the outer branches of 112 $P_{iso}(E)$ plots with 250 values of $E$. The map of $T_{adi}(P,E)$ (Fig. 1e) shows the outer branches of 358 $P_{adi}(E)$ plots with 250 values of $E$.

**(a) Interpolation.** To obtain |Δ$T$| while following a particular plot of $P_{adi}(E)$ from 0 to $E$, we identify the start temperature by interpolating between the temperatures of the two $P_{iso}(E)$ plots that lie either side, and we identify the finish temperature likewise.

**(b) Averaging.** For the maximum field change of 0→26 kV cm$^{-1}$, we can find |Δ$T$| in this way by using the first data point ($j = 1$) and the last data point ($j = 250$) in the chosen plot of $P_{adi}(E)$. In view of noise, we averaged 15 values of |Δ$T$| that were evaluated for 15 similar changes of field, using the second data point ($j = 2$) with the second last data point ($j = 249$), the third data point ($j = 3$) with the third last data point ($j = 248$), and so on, up to the fifteenth data point ($j = 15$) with the fifteenth last data point ($j = 236$). The field change for the thus averaged value of |Δ$T$| may therefore be regarded as 0.75±0.75 kV cm$^{-1}$ → 25.25±0.75 kV cm$^{-1}$.

For temperatures below $T_C \sim$ 295 K or above 315 K, the values of |Δ$T$| are spuriously large, as the high density of $P_{iso}(E)$ and $P_{adi}(E)$ contours in the $T_{iso}(P,E)$ and $T_{adi}(P,E)$ maps implies that any random or systematic errors will lead to contour mismatch.

## Note 2. Evaluation of electrocaloric $|\Delta T|$ from $\chi^2$

The $\chi^2$ approach described below represents a slightly less direct route to $|\Delta T|$, but enables a more formal error analysis. To compare two specific adiabatic and isothermal $P(E \geq 0)$ branches with each other, we define test statistic $\chi^2$ as the sum of the squares of their difference in $P$:

$$\chi^2 = \sum_{j=1}^{N} \left[P_{\text{adi}}(E_j) - P_{\text{iso}}(E_j)\right]^2 / \sigma^2$$

where $E_j$ is the field corresponding to the $j$th discrete datapoint in the $P(E)$ branch (with a total of $N = 250$ datapoints), , and we assume that individual measurements of $P$ follow a Gaussian distribution with standard deviation $\sigma$.

Given that $\sigma$ is not known initially, we evaluate $\sigma^2 \chi^2 = \sum_{j=1}^{N} \left[P_{\text{adi}}(E_j) - P_{\text{iso}}(E_j)\right]^2$ for all 40096 combinations of the 112 $P_{\text{iso}}(E)$ and 358 $P_{\text{adi}}(E)$. Fig. S1(a) (left vertical axis) shows a representative subset of the results of this calculation, for four of the 358 $P_{\text{adi}}(E)$, and all of the 112 $P_{\text{iso}}(E)$ (horizontal axis). The dominant feature of each curve is a single inverted peak, whose minimum identifies the $P_{\text{iso}}(E)$ that most closely matches the $P_{\text{adi}}(E)$. By identifying this $\sigma^2 \chi^2$ minimum (Fig. S1(b), left vertical axis) in the 112 $P_{\text{iso}}(E)$, for each of the 358 $P_{\text{adi}}(E)$ (Fig. S1(b), horizontal axis), we observe that the values of such minima at high temperatures exceed those values at low temperatures by orders of magnitude. This is because the high-temperature $P_{\text{iso}}(E)$ and $P_{\text{adi}}(E)$ data sample distributions are different due to the presence of large EC effects, whereas the low-temperature $P_{\text{iso}}(E)$ and $P_{\text{adi}}(E)$ data sample the same distribution, as there is no significant EC effect. In this low-temperature regime, $\sigma^2 \chi^2$ is close to $\sigma^2 N$, given that $P_{\text{adi}}$ and $P_{\text{iso}}$ are taken from the same distribution, such that $|P_{\text{adi}}(E)-P_{\text{iso}}(E)|$ averaged over the 250 datapoints is close to $\sigma$. Thus, by considering the value of the $\sigma^2 \chi^2$ at temperatures $T < 290$ K, we make a conservative estimate of $\sigma = 0.14$ µC cm$^{-2}$ (dotted line, Fig. S1(b)). With the value of $\sigma$ known, we now map all values of $\sigma^2 \chi^2$ plotted in Fig. S1(a-b) (left vertical axes) to values of $\chi^2$ (right vertical axes).

Knowledge of the $\chi^2$ minimum (Fig. S1(b)) is sufficient to directly compute the p-value (Fig. 3(c)) via a standard $\chi^2$ distribution, which represents the confidence level with which our data support the presence of an EC effect. As expected, p-values are close to 0 at $T < 290$ K, and close to 1 at higher temperatures.

The $\chi^2$ minimum (Fig. S1(b)) is not sufficient to evaluate EC $|\Delta T(T)|$, and it is necessary to consider the 40096 values of $\chi^2$ from which the $\chi^2$ minimum is derived (Fig. S1(a)). By considering $\chi^2$ as a function of the 112 $P_{iso}(E)$ [Fig. S1(a)], it is possible to relate $|\Delta T|$ to the width of the inverted peak that defines the $\chi^2$ minimum, by a straightforward geometric argument, as explained below.

For a field change of $E_1 \rightarrow E_2$, the EC effect may take place continuously between those fields, or abruptly at some critical field $>E_1$ and $<E_2$, or in some other way whose fashion lies intermediate between these two extremes. Fig. S2(a-b) illustrate the continuous and abrupt cases respectively, in terms of how the adiabat (red line) intersects the isotherms (black lines). The isotherm $T$ corresponds to the $\chi^2$ minimum, whereas the isotherms $\pm\frac{1}{2}\Delta T$ correspond to the boundaries spanned by the adiabatic data.

Recalling that $\chi^2$ is the integrated squared deviation of an adiabat from an isotherm, we illustrate this deviation for isotherm $T$ in Fig. S2(c-d) and isotherms $\pm\frac{1}{2}\Delta T$ in Fig. S2(e-f), for continuous and abrupt EC effects. A straightforward calculus shows that for continuous EC effects, $\chi^2$ for the isotherms $\pm\frac{1}{2}\Delta T$ exceeds that for isotherm $T$ by a factor of four. For abrupt EC effects, $\chi^2$ for isotherms $\pm\frac{1}{2}\Delta T$ exceeds that for isotherm $T$ by a factor of two. Given that the character of experimental data must lie between these two limiting cases, the full width-at-double minimum (FWDM) and the full width-at-quadruple minimum (FWQM) of the inverted peak of $\chi^2$ represent lower and upper bounds of EC $|\Delta T|$, respectively.

We plot the FWDM and the FWQM for the 358 $P_{adi}(E)$ in Fig. S3. Given that aspects of both continuous and abrupt character are readily apparent in our $P(E)$ data (Fig. 2 of main text), with approximately ¼ of the $E$-axis span occupied by the rate of maximum change, we adopt a weighted average of ¼ FWDM and ¾ FWQM for our plot of EC $|\Delta T|$.

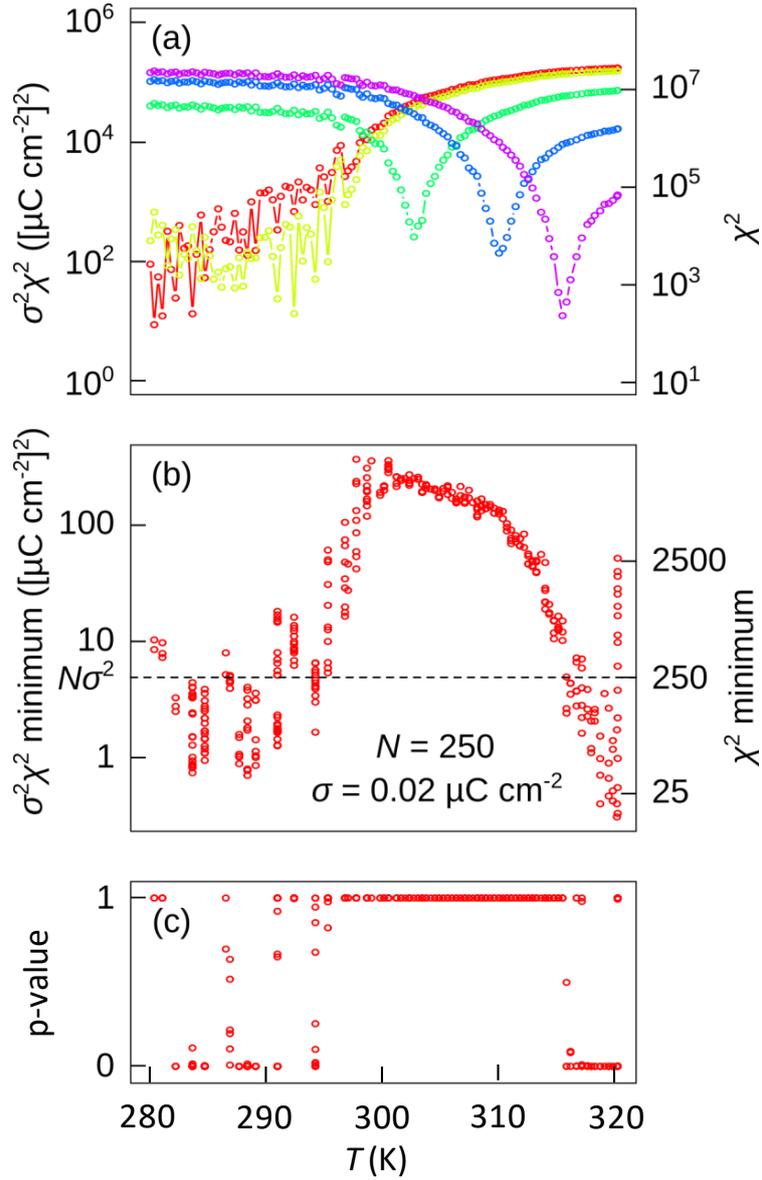

Figure S1. Statistical analysis of the 112 isothermal and 358 adiabatic $P(E\geq 0)$ outer branches. (a) For four of the 358 adiabatic $P_{adi}(E\geq 0)$, we present $\sigma^2\chi^2$ by comparison with the 112 isothermal $P_{iso}(E\geq 0)$. Each datapoint of each coloured dataset is from one of the 112 isothermal plots. (b) For all of the 358 $P_{adi}(E\geq 0)$, the $\sigma^2\chi^2$ minimum (on scanning through the 112 isothermal plots) is plotted. This enables $\sigma^2$ to be identified (dotted line) and the data in (a) and (b) to be mapped to values of $\chi^2$ itself (right vertical axes). (c) p-value for the $\chi^2$ minima in (b).

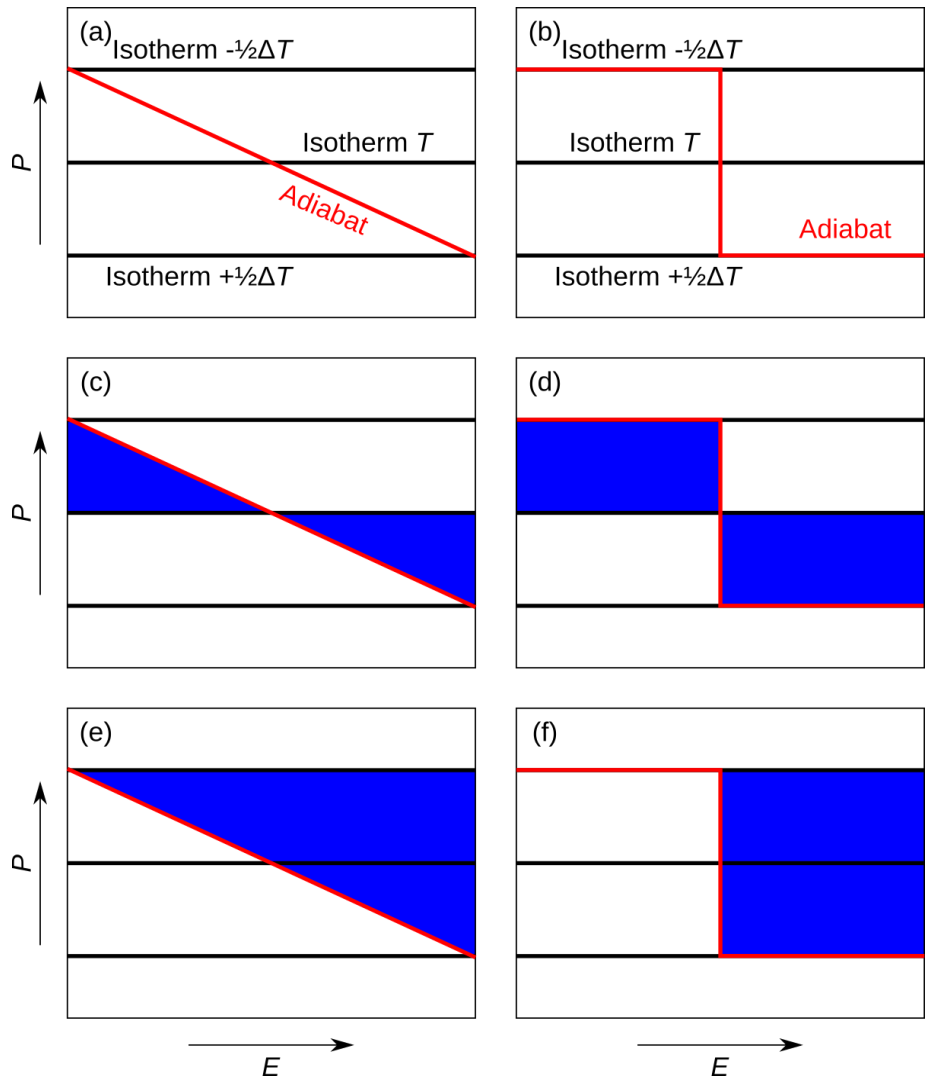

Figure S2. (a-b) Toy illustration of the intersection between an adiabat and three isotherms. Isotherm $T$ corresponds to the $\chi^2$ minimum. Panel (a) is for a continuously evolving EC effect, whereas panel (b) is for an abrupt EC effect, with realistic behaviour lying between the two cases. (c-d) Illustration of the $P$-$E$ space (blue area) subject to integrated squared deviation, for the evaluation of $\chi^2$ for isotherm $T$. (e-f) As for (c-d), but for isotherm $\pm\tfrac{1}{2}\Delta T$.

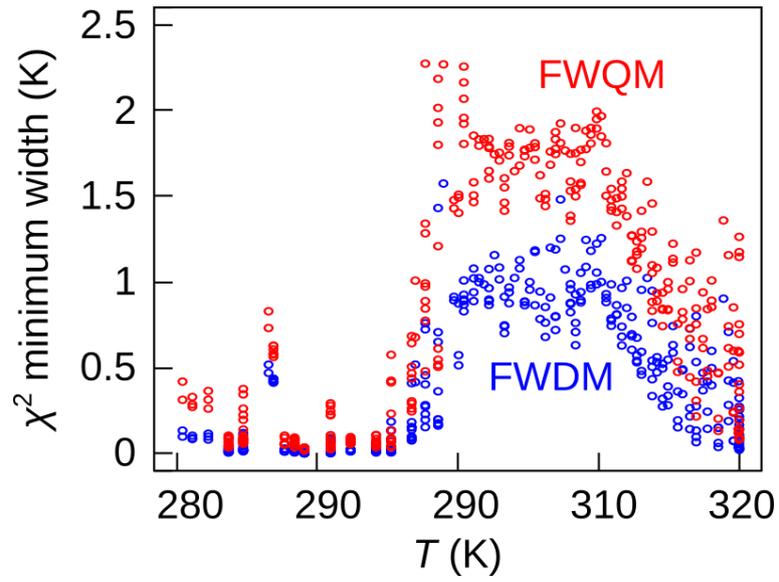

Figure S3. The full width-at-double minimum (FWDM) and the full width-at-quadruple minimum (FWQM) of the inverted peak of $\chi^2$ (Fig. S3(a)). Each datapoint corresponds to one of the 358 $P_{\text{adi}}(E)$.